\documentclass[longauth]{aa} 
\usepackage{graphicx}
\usepackage{txfonts}
\usepackage{textcomp}
\usepackage{color}
\usepackage[nointegrals]{wasysym}
\usepackage{amssymb}
\usepackage[separate-uncertainty = true,multi-part-units=single]{siunitx}
\usepackage{babel}
\usepackage{hyperref}
\usepackage[switch]{lineno}

\bibpunct{(}{)}{;}{a}{}{,}

\PassOptionsToPackage{american}{babel}
\PassOptionsToPackage{authoryear}{natbib}

\renewcommand{\deg}{^\circ}

\usepackage[normalem]{ulem}

\usepackage{scrextend}

\begin{document} 

\title{MAGIC very large zenith angle observations of the Crab Nebula up to 100~TeV}


%
%
\author{
MAGIC Collaboration:
V.~A.~Acciari\inst{1} \and
S.~Ansoldi\inst{2,23} \and
L.~A.~Antonelli\inst{3} \and
A.~Arbet Engels\inst{4} \and
D.~Baack\inst{5} \and
A.~Babi\'c\inst{6} \and
B.~Banerjee\inst{7} \and
U.~Barres de Almeida\inst{8} \and
J.~A.~Barrio\inst{9} \and
J.~Becerra Gonz\'alez\inst{1}
% ...
}
\institute { Inst. de Astrof\'isica de Canarias, E-38200 La Laguna, and Universidad de La Laguna, Dpto. Astrof\'isica, E-38206 La Laguna, Tenerife, Spain
\and Universit\`a di Udine, and INFN Trieste, I-33100 Udine, Italy
\and National Institute for Astrophysics (INAF), I-00136 Rome, Italy
\and ETH Zurich, CH-8093 Zurich, Switzerland
\and Technische Universit\"at Dortmund, D-44221 Dortmund, Germany
\and Croatian Consortium: University of Rijeka, Department of Physics, 51000 Rijeka; University of Split - FESB, 21000 Split; University of Zagreb - FER, 10000 Zagreb; University of Osijek, 31000 Osijek; Rudjer Boskovic Institute, 10000 Zagreb, Croatia
\and Saha Institute of Nuclear Physics, HBNI, 1/AF Bidhannagar, Salt Lake, Sector-1, Kolkata 700064, India
\and Centro Brasileiro de Pesquisas F\'isicas (CBPF), 22290-180 URCA, Rio de Janeiro (RJ), Brasil
\and IPARCOS Institute and EMFTEL Department, Universidad Complutense de Madrid, E-28040 Madrid, Spain
\and University of \L\'od\'z, Department of Astrophysics, PL-90236 \L\'od\'z, Poland
\and Universit\`a di Siena and INFN Pisa, I-53100 Siena, Italy
\and Deutsches Elektronen-Synchrotron (DESY), D-15738 Zeuthen, Germany
\and Istituto Nazionale Fisica Nucleare (INFN), 00044 Frascati (Roma) Italy
\and Max-Planck-Institut f\"ur Physik, D-80805 M\"unchen, Germany
\and Institut de F\'isica d'Altes Energies (IFAE), The Barcelona Institute of Science and Technology (BIST), E-08193 Bellaterra (Barcelona), Spain
\and Universit\`a di Padova and INFN, I-35131 Padova, Italy
\and Universit\`a di Pisa, and INFN Pisa, I-56126 Pisa, Italy
\and ICRANet-Armenia at NAS RA, 0019 Yerevan, Armenia
\and Centro de Investigaciones Energ\'eticas, Medioambientales y Tecnol\'ogicas, E-28040 Madrid, Spain
\and Universit\"at W\"urzburg, D-97074 W\"urzburg, Germany
\and Finnish MAGIC Consortium: Finnish Centre of Astronomy with ESO (FINCA), University of Turku, FI-20014 Turku, Finland; Astronomy Research Unit, University of Oulu, FI-90014 Oulu, Finland
\and Departament de F\'isica, and CERES-IEEC, Universitat Aut\`onoma de Barcelona, E-08193 Bellaterra, Spain
\and Japanese MAGIC Consortium: ICRR, The University of Tokyo, 277-8582 Chiba, Japan; Department of Physics, Kyoto University, 606-8502 Kyoto, Japan; Tokai University, 259-1292 Kanagawa, Japan; RIKEN, 351-0198 Saitama, Japan
% ...
}

\date{Received XX XX, 2016; accepted XX XX, 2016}


 
  \abstract
   {}  
   {We aim to measure the Crab Nebula $\gamma$-ray spectral energy distribution in the $\sim 100$~TeV energy domain and test the validity of existing leptonic emission models at these high energies.}
   {We use the novel very large zenith angle observations with the MAGIC telescope system to increase the collection area above 10~TeV. We also develop an auxiliary procedure of monitoring atmospheric transmission in order to assure proper calibration of the accumulated data. This employs recording of optical images of the stellar field next to the source position, which provides a better than 10\% accuracy for the transmission measurements.}
   {We demonstrate that MAGIC very large zenith angle observations yield a collection area larger than a square kilometer. In only $\sim 56$~hr of observations, we detect the $\gamma$-ray emission from the Crab~Nebula up to 100~TeV, thus providing the highest energy measurement of this source to date with Imaging Atmospheric Cherenkov Telescopes. Comparing accumulated and archival MAGIC and \textit{Fermi}/LAT data with some of the existing emission models, we find that none of them provides an accurate description of the 1~GeV to 100~TeV $\gamma$-ray signal.}
   {}

  
  \titlerunning{MAGIC VLZA observations of the Crab Nebula up to 100~TeV}
  \maketitle


\section{Introduction}

The Crab Nebula broad-band emission is usually interpreted in the framework of leptonic models. The radio to MeV gamma-ray emission is attributed to synchrotron radiation of energetic electrons in the $120-150~\mu G$ nebula magnetic field. At higher energies, GeV to TeV emission is linked to the inverse Compton (IC) scattering on the synchrotron, infra-red (IR) and Cosmic Microwave (CMB) background photons. The morphology of the Nebula, revealed by the optical and X-ray data, is non-trivial. Nonetheless, its broadband spectral energy distribution (SED) is reasonably well described even by one-zone models, involving diverse electron distributions~\citep{Meyer_Crab} and/or propagation/cooling effects~\citep{Martin_Crab, Fraschetti_Pohl_Crab}.

At the same time, most of the proposed models fail to describe the details of the Crab Nebula SED~\citep{MAGIC_Crab15}. The highest model-to-data deviations lie in the keV to MeV range, where the SED softens, and GeV to TeV range, where the IC~peak appears broader than suggested by several models. In addition to IC-related emission, bremsstrahlung and proton-proton interactions may also contribute to the GeV-TeV emission, if emitting electrons are at least partially confined in the filaments of the nebula, filled with the ionised gas~\citep[][see also Sect.~\ref{sect::summary}]{atoyan96}.

A way to resolve this degeneracy is offered by observations at the highest energies above several tens of TeV, where at least the bremsstrahlung process gives subdominant contribution to the Nebula emission. The dominant emission at those energies is due to the combination of the synchrotron-self-compton (SSC) and IC/CMB emission of electrons with energies $\gtrsim 10^{13}$~eV. The SSC part of the emission at these energies is produced in the deep Klein-Nishina regime $E^\gamma_{bkg} E_e / (m_e^2 c^4) \gtrsim 1$ with $E^\gamma_{bkg}$ being the energy of the background photons, $E_e$ -- that of electrons, $m_e$ the electron mass and $c$ the speed of light). Due to this the SSC spectrum traces that of the underlying electron population. This way, the apparent changes in the synchrotron spectrum at keV-MeV energies should also manifest themselves in the $\sim 10-100$~TeV energy band. The absence of the corresponding spectral changes at these energies would indicate the sub-dominant nature of the leptonic SSC emission at the highest energies, emitted by the Nebula -- in favour of other competing mechanisms.

Observations at energies $\gtrsim 10$~TeV are usually associated with low event count rates from astrophysical sources. The collection area $A_{eff}$ of Imaging Atmospheric Cherenkov Telescopes (IACTs) is determined by the size of the Cherenkov light cone from the $\gamma$-ray induced extended air showers (EAS). For vertical observations the collection area of a single telescope is $\sim 0.05~\mathrm{km}^2$. The collection area can be increased using a larger number of telescopes, like in the forthcoming Cherenkov Telescope Array (CTA) observatory. Alternatively, a similar effect can be achieved by using observations at higher zenith angles~\citep{Sommers_VLZA_1987} \citep[see also][]{Konopelko_LZA, MAGIC_GC}. This observation mode leads to an increase in the Cherenkov pool size due to the larger distances to the showers. At the same time the reduced photon density on the ground shifts the energy threshold of the telescope to significantly higher energies. Technical details of the novel VLZA observation technique can be found in~\citet{VLZA_NIM}.

In this paper we present the results of the Crab Nebula observation at very large zenith angles (VLZA; $>70\deg$) with the MAGIC telescopes and discuss them in the context of other multi-wavelength data of this source.

\section{MAGIC very large zenith angle observations of the Crab Nebula}
\label{sect::magic_observations_general}


\subsection{The MAGIC Telescopes}

The MAGIC (Major Atmospheric Gamma Imaging Cherenkov) telescopes are a system of two 17~m diameter IACTs, located at an altitude of 2200~m a.s.l. at the Roque de los Muchachos Observatory on the Canary Island of La Palma, Spain (28$^\circ$N, 18$^\circ$W).

The telescopes are used to image flashes of Cherenkov light produced by the charged component of EAS initiated in the upper atmosphere by gamma-ray photons with energies $\gtrsim 30$~GeV. Both telescopes are nominally operated together in a coincidence (so-called stereoscopic) mode, in which only events simultaneously triggering both telescopes are recorded and analyzed~\citep{Magic_performanceII}. For low zenith angle (ZA; $<30\deg$) observations and for \mbox{$E>220$~GeV}, the integral sensitivity of MAGIC is $(0.66\pm0.03)\%$ in units of the Crab Nebula flux (C.U.) for 50 hours of observations~\citep{Magic_performanceII}. 


\subsection{Observations}
\label{sect::observations}

The data sample presented here was accumulated from December 2014 until November 2018 in the zenith angle range $70\deg-80\deg$ and comprises of $\approx 56$~hr of good-quality data (after the initial data selection as described below; $\approx 88$~hr before the selection), taken in the so-called ``wobble'' mode~\citep{1994Fomin_wobble}. The summary of observational time per year is given in Table.~\ref{tab::obs_summary}.

VLZA observations of the Crab Nebula by MAGIC can be performed in two configurations -- during the source rise or set above the horizon. These configurations give somewhat different sensitivities of the MAGIC stereoscopic system to the incoming $\gamma$-ray flux due to the varying stereo baseline -- projected inter-telescope distance seen from the direction of the source. The two MAGIC telescopes are located northeast and southwest from the system center, thus providing a larger baseline in the North-West-North and South-East-South directions. With the declination of $\approx 22\deg$, Crab Nebula rises at $\approx 77\deg$ and sets at $\approx 283\deg$ azimuths (counting from the North), i.e. North-East-North and North-West-North correspondingly. Here both ``rise'' and ``set'' configurations were used with most of the data taken in the ``set'' direction due to the larger stereo baseline.

Observations in the VLZA regime correspond to shower distances of $\gtrsim 50-100$~km from the telescopes as opposed to $\sim 10$~km at lower zenith angles $\lesssim 30\deg$. As such, these measurements are subject to increased light attenuation due to the scattering and absorption in the atmosphere. The standard MAGIC way to account for such effect -- utilisation of a dedicated micro-LIDAR system~\citep{fruck_novel_2014} -- allows only to probe the atmospheric absorption at distances $\lesssim 20$~km. To ensure an appropriate control over the wavelength-dependent atmosphere attenuation we took additional contemporaneous images of the stellar field next to the Crab Nebula with red, green and blue filters which allows to monitor the total atmospheric transmission with an accuracy better than 10\%. The details of this procedure are given in Appendix~\ref{sect::sbig_correction}; see also~\citet{VLZA_NIM}.

We did not change the optical focusing of the telescopes (usually set to the 10~km distance), as our tests with Monte Carlo simulation did not indicate any significant performance improvement from doing so.

\begin{table}
  \centering
  \caption{Summary of the duration of the MAGIC Crab Nebula VLZA observations. The observational time is given separately for the rise and set of the source on the horizon.}
  \begin{tabular}{l | r r r r r}
    \hline\hline
    Year          &  2014 & 2015 & 2016 & 2017 & 2018 \\
    \hline
    Target rising [h]  &   0.0 & 2.43 &  6.43 &  7.03 &  3.90 \\
    Target setting [h] &  1.17 & 4.00 & 10.37 & 16.46 &  4.44 \\
    \hline
    Total [h]          &  1.17 & 6.43 & 16.80 & 23.50 &  8.34 \\
    \hline
  \end{tabular}
  \tablefoot{
    Observation time is given in hours, after the data selection cuts.
  }
  \label{tab::obs_summary}
\end{table}


\section{Data analysis}
\label{sect::data_analysis_general}


\subsection{MAGIC data analysis}
\label{sect::magic_analysis}

The acquired data are reduced with the standard MAGIC Analysis and Reconstruction Software~\citep[MARS;][]{zanin2013mars}. We first remove events detected during adverse weather conditions and those corresponding to the known temporary hardware issues. Due to the nature of VLZA observations several usual data cuts are no longer efficient. These include the presence of the clouds in the telescope's field of view, measured with an infra-red pyrometer system~\citep{gaug_atmospheric_2014}, and the number of stars detected by the MAGIC star-guider cameras. The corresponding measurements were used to cross-check the applied event selection. The latter was based on the cuts on the mean currents of photomultipliers, the event trigger rate and the LIDAR transmission at the maximal accessible range.

We use the standard MAGIC MARS routines to reconstruct the initial direction and impact distance with respect to the MAGIC telescopes for the recorded EAS images. These were augmented with contemporaneous atmospheric transmission monitoring and corrections, as explained in Appendix~\ref{sect::sbig_correction}.

To reconstruct the energy of the EAS initiating particle, three different methods were used: (a) standard MARS procedure based on a look-up table~\citep[LUT,][]{2012MAGICPerformance} created from the MAGIC Monte Carlo (MC) simulations, (b) random forest (RF) multivariate analysis and (c) neural network (NN) regression. The applied cosmic-ray background suppression was based on the classification scheme implemented with both RF~\citep{Magic_RF, 2012MAGICPerformance} and NN. All these techniques yield consistent results. To derive the results shown below, LUT energy estimation and RF event classification techniques were employed.

\subsection{\textit{Fermi}/LAT data analysis}

In this work we made use of the publicly available \textit{Fermi}/LAT Pass~8 data set\footnote{http://fermi.gsfc.nasa.gov/cgi-bin/ssc/LAT/LATDataQuery.cgi}. We have used the \textit{Fermi Science Tools} package\footnote{http://fermi.gsfc.nasa.gov/ssc/data/analysis/scitools/} v9r33p0 for data processing, retaining only the ``Source'' (P8R2\_SOURCE\_V6) class events, registered till June 2017 within the $75^\circ$ zenith angle; we have not applied the ROI-based zenith angle cut. The photons selected from the $20^\circ$ region around the Crab Nebula position were further required to lie within the Crab Pulsar 0.60-0.82 phase range where the nebula emission dominates~\citep{fermi_crab}. We have not applied an additional gating of the Crab Nebula flares~\citep{Crab_flare_2009, Crab_flare_2011}. These flares mostly affect the low-energy synchrotron emission of the Nebula and due to their $\lesssim 500$~MeV cut-off energy are practically undetectable above $\sim 3$~GeV energy, and even in the $0.3-1$~GeV energy range contributions of the flares to the average Nebula flux in 9~years is limited to $5-10\%$ due to their short duration. The fluxes of all the sources in the 300~MeV -- 510~GeV energy range in the selected region were estimated from the joint likelihood fit. Given the Crab Nebula brightness in the \textit{Fermi}/LAT energy range, the fitted model included the diffuse background components (namely ``iso\_P8R2\_SOURCE\_V6\_v06.txt'' and ``gll\_iem\_v06.fits'') and 20 brightest objects from the 3FGL~\citep{3FGL} catalogue within 28 degrees from the source of interest. The Crab Nebula model itself comprised a single power law; employment of more complex multi-component models is not required with the narrow energy bins (5 per decade) used here.


\section{Results}
\label{sect::results}

During the VLZA data taking, the low energy threshold after the data selection cuts quickly increases from $\sim 1$~TeV at zenith angle of $70\deg$ to $\sim 10$~TeV when approaching $80\deg$. The collection area at energies above 70~TeV quickly reaches approximately 2~km\textsuperscript{2} \citep[compared to $\sim 0.1$~km\textsuperscript{2} for low zenith angle observations,][]{Magic_performanceII}, leading to an unprecedented gamma-ray collection area.

\begin{figure}
  \includegraphics[width=\columnwidth]{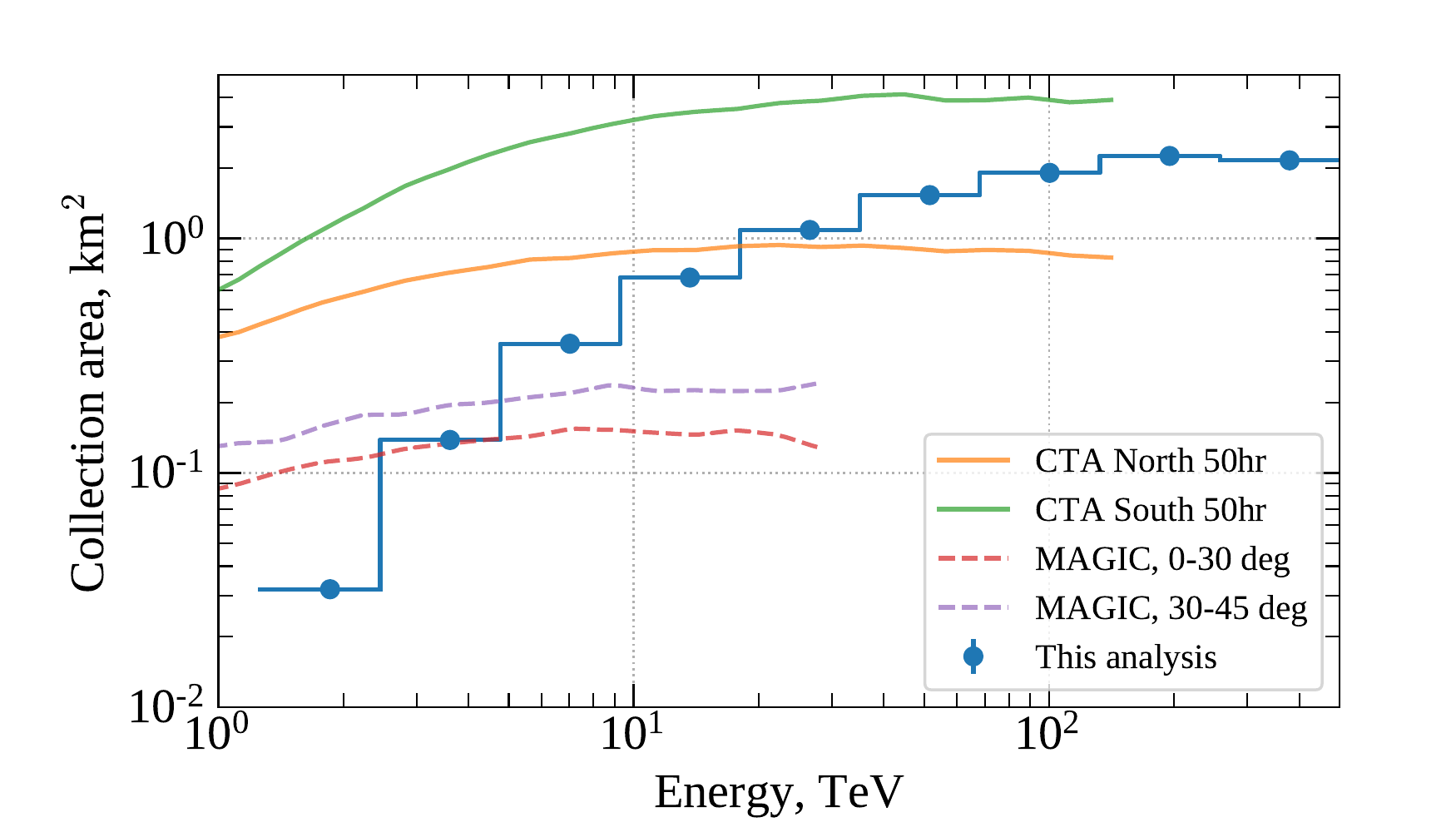}
  \caption{MAGIC collection area, estimated for an observational sample in the zenith angle range $70^\circ-80^\circ$ with Monte Carlo simulations~\citep[see also][]{VLZA_NIM}. For comparison, the collection area (for $20^\circ$ zenith angle observations in so-called ``Production 3'' layout) of the future CTA array is also shown (see Sect.~\ref{sect::results} for details). Also shown is the MAGIC collection area at lower zenith angles from~\citet{Magic_performanceII}.}
  \label{fig::vlza_aeff}
\end{figure}

To estimate the MAGIC performance for the acquired VLZA data, we have used a dedicated Monte Carlo simulation, describing the MAGIC observations of gamma-ray induced air showers in the zenith angle range $70^\circ-80^\circ$. This simulation was performed with the Corsika code~\citep{Corsika}, modified to include the MAGIC specific output. It also included the curvature of the Earth's atmosphere to properly describe the increasing air column density during the near-horizon observations. The rest of the simulation procedure was performed the same way as for the lower zenith angle observations~\cite[e.g.][]{Magic_performanceII}. The resulting collection area estimated after the data selection cuts is shown in Fig.~\ref{fig::vlza_aeff}. For comparison, the expected collection area of the currently under construction CTA\footnote{Expected CTA performance can be found  here: \url{https://www.cta-observatory.org/science/cta-performance/}} is shown.

It should be noted, that despite of the increase of the collection area, MAGIC VLZA performance is impacted by the limited reconstruction of the shower parameters, resulting from their remoteness ($\gtrsim 50-100$~km) and correspondingly smaller image size in the telescope camera. In case of MAGIC, the measured images sizes are decreased down to 3-4 pixels for most of the detected showers. Though the number of excess events in our data sample changes with energy as expected given the collection area, we find that these small images degrade the performance of the cosmic-ray background suppression technique we employ. In addition we also note a $\sim 1.5-2$~fold degradation of the energy and angular resolution, compared to observations at small zenith angles~\citep{Magic_performanceII}. Given that this is a novel technique for IACT observations, we anticipate that a certain improvements can be achieved with dedicated, optimized analysis. Still we find that processing of the VLZA data with the standard MAGIC MARS tools at present allows us to perform the shown below interesting studies.

In 56~hours of observations, the Crab Nebula signal at estimated energies above 30~TeV was detected at a $\approx 6.5 \sigma$~\citet{li_analysis_1983} significance level. Despite the increased energy threshold, the spectrum could be reconstructed down to the energy of $\sim 1$~TeV.

In order to reconstruct the Crab Nebula SED at energies above 1~TeV, we have made use of all three energy estimation methods outlined in Sect.~\ref{sect::magic_analysis}. We applied the background rejection with both the standard MARS routines and dedicated NN, each time adjusting the cuts so as to maintain $90\%$ of Monte Carlo gamma rays in each energy bin. 

\begin{figure}
  \includegraphics[width=\columnwidth]{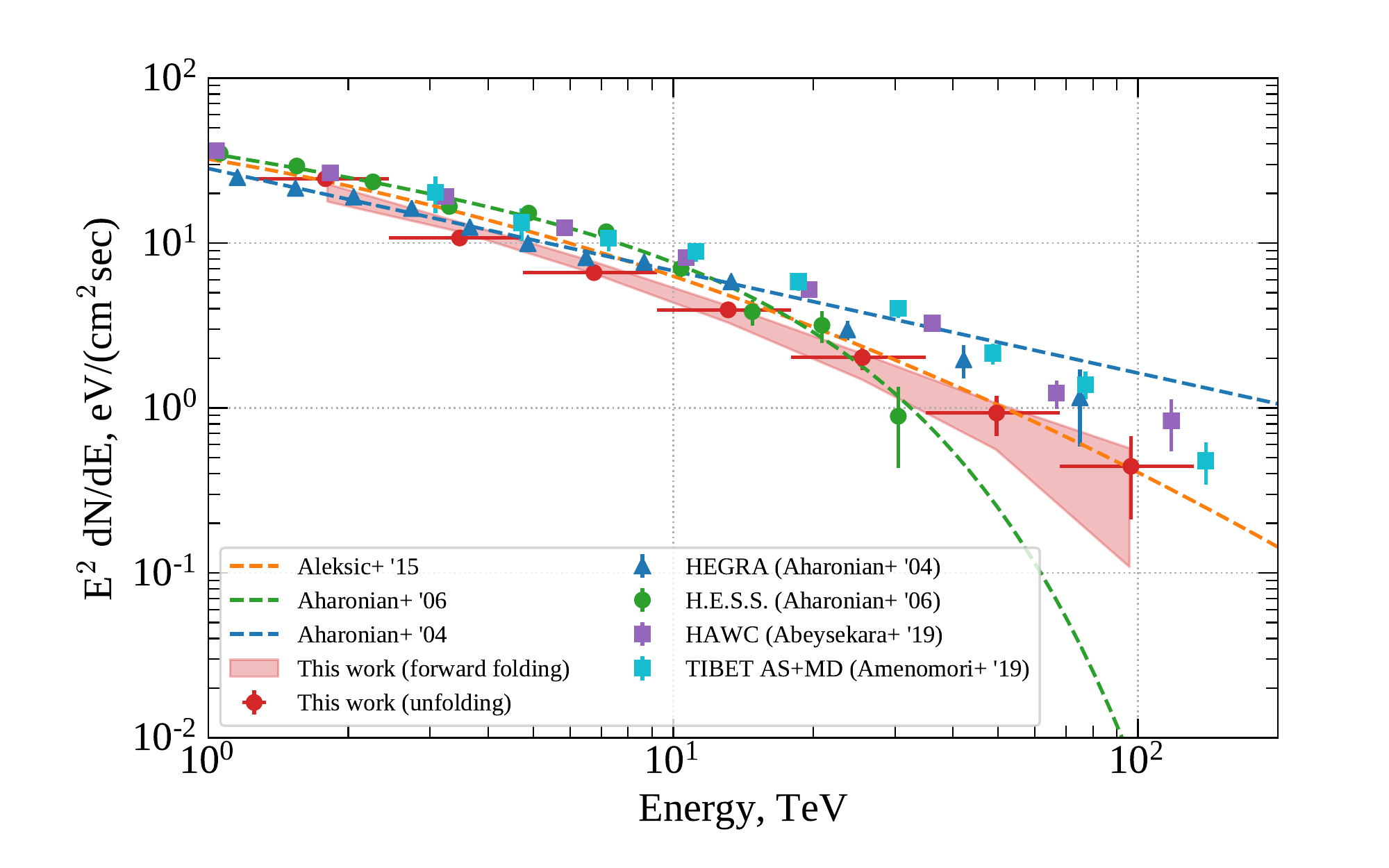}
  \caption{Spectral energy distribution of the Crab Nebula, obtained with MAGIC VLZA observations (red). Both results of the spectral unfolding (data points) and forward folding (band) procedures are shown. Dashed colored lines denote the archival best fits to the Crab Nebula spectra from~\citet{HEGRA_Crab}, \citet{HESS_Crab} and~\citet{MAGIC_Crab15} correspondingly. Data points from~\citet{HEGRA_Crab}, \citet{HESS_Crab}, \citet{HAWC_Crab_100TeV} and \citet{Tibet_Crab_100TeV} are also shown for comparison.}
  \label{fig::vlza_crab_spectrum}
\end{figure}

\begin{figure}
  \includegraphics[width=\columnwidth]{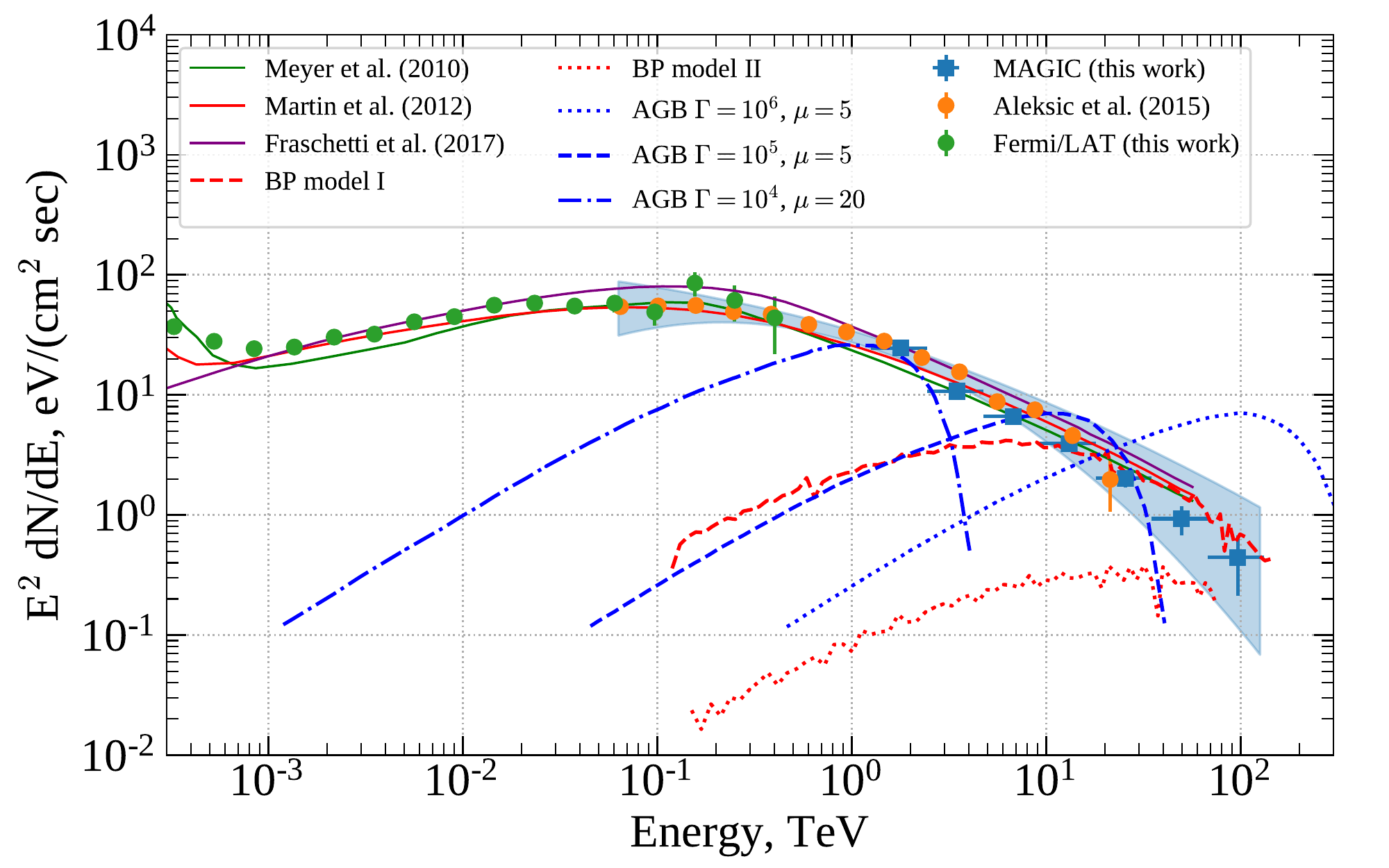}
  \caption{
    Crab Nebula spectrum obtained here compared to the lower energy measurements with MAGIC~\citet{MAGIC_Crab15} and \textit{Fermi}/LAT. The blue shaded band denotes the 68\% confidence flux range, estimated from the fit with the log-parabola function. 
    Solid lines of different colours show several leptonic models from~\citet{Meyer_Crab, MAGIC_Crab15, Fraschetti_Pohl_Crab}, previously constructed for the Crab Nebula.
    Dashed lines denote predictions for hadronic contribution from~\citet{bednarek97}, referred to as ``BP'', and~\citet{amato03}, marked as ``AGB''. $\Gamma$ denotes the bulk wind Lorentz factor, whereas $\mu=n/\overline{n}$ is the effective target material density increase over its mean value.
    }
  \label{fig::GeV-TeV_SED}
\end{figure}

The SED of the Crab Nebula up to $\sim100$~TeV, obtained with the LUT energy estimation method, is shown in Fig.~\ref{fig::vlza_crab_spectrum}. 
It can be seen from Fig.~\ref{fig::vlza_crab_spectrum} that the previous HEGRA~\citep{HEGRA_Crab} spectrum, produced with about 400~hr of data, is within $\lesssim 20\%$ from the MAGIC results. A comparison of these data with the previous lower energy measurements -- including the lower zenith angle MAGIC observations from~\citet{MAGIC_Crab15} -- is given in Fig.~\ref{fig::GeV-TeV_SED}. Our data do not support the indications for the $\sim 30$~TeV high-energy cut-off, suggested earlier~\citep{HESS_Crab}.

In figure~\ref{fig::GeV-TeV_SED} we also fit the Crab Nebula SED above 60~GeV with the log-parabola function $dN/dE = f_{0} (E/E_{0})^{(\alpha + \beta \log_{10}(E/E_{0}))}$, also using the archival MAGIC data~\citep{MAGIC_Crab15} in addition to the VLZA measurements. Accounting for 15\% (17\% in VLZA case) systematics in the MAGIC energy scale and 11\% (20\% in VLZA case) on the flux scale~\citep{MAGIC_Crab15, Magic_performanceII}, we find this fit to be in poor agreement with the data ($\chi^2 \approx 33$ over 14 degrees of freedom), indicating that the log-parabola does not provide a good match to the Crab Nebula spectral shape over the entire $60$~GeV -- $100$~TeV energy range. Otherwise the best-fit parameters are similar to those found in~\citep{MAGIC_Crab15}: for fixed $E_0=1$~TeV we find $\alpha=-2.48 \pm 0.03$, $\beta=-0.23 \pm 0.01$ and $f_0 = (2.95 \pm 0.27) \times 10^{-23}~\mathrm{ph/(cm^2~sec~eV)}$ (all uncertainties correspond to 1$\sigma$ confidence range).

As one can see from Fig.~\ref{fig::GeV-TeV_SED}, the overall shape of the GeV to 100~TeV $\gamma$-ray emission of the Crab Nebula can be reasonably described within the framework of existing theoretical models like \citet{Meyer_Crab}, \citet{Martin_Crab} and \citet{Fraschetti_Pohl_Crab}. At the same time, these models do not reproduce the gradual softening of the Inverse-Compton (IC) emission peak at multi-TeV energies. A simultaneous fit of \textit{Fermi}/LAT, archival and VLZA MAGIC data to the best-fit model curves yields a relatively large $\chi^2$ values -- $183.0/26$~d.o.f. for \citet{Meyer_Crab}, $77.5/26$~d.o.f. for \citet{Martin_Crab} and $140.0/26$~d.o.f. for \citet{Fraschetti_Pohl_Crab} (as presented in \citet{MAGIC_Crab15}). It should be noted though, that these large $\chi^2$ values seem to be dominated by the point-to-point systematics, not accounted for here.


\section{Discussion}
\label{sect::summary}

The novel method of VLZA observations with the MAGIC telescopes allows one to detect gamma rays up to hundreds of TeV in a few tens of hours. It can be efficiently used to search for astrophysical sources accelerating particles to PeV energies. This observation technique requires simultaneous measurements of the atmospheric transparency, careful studies of systematics and properly tailored MC data. To this extent we have developed an auxiliary atmospheric transmission measurement procedure employing contemporaneous measurements of stellar light from the region next to the $\gamma$-ray target.

As already discussed earlier, the remoteness of the showers from the telescope during VLZA observations has an impact on the instrument performance. At present this impacts our ability to infer the highest energy flux from the soft-spectrum sources like Crab Nebula. Along with other differences in the VLZA shower development~\citep{neronov16}, this indicates that a revision of the analysis technique is needed to fully unveil the full potential of VLZA data taking.

Still, the obtained VLZA Crab Nebula data at few tens of TeV agree well with the earlier, lower zenith angle measurements with HEGRA~\citep{HEGRA_Crab}, that were obtained in $\sim 8$~times longer observational time. VLZA measurements presented here also support the source spectrum extension up to 100~TeV and likely beyond. These results are consistent with findings of HAWC~\citep{HAWC_Crab_100TeV} and Tibet AS$\gamma$ collaborations~\citep{Tibet_Crab_100TeV}.

Generally, the available multi-wavelength Crab Nebula flux measurements can be explained within the framework of leptonic models. The latter, however, do not provide much flexibility in the spectral shape of the inverse Compton emission part~\citep{atoyan96}. Testing the source flux ratios in the energy bins above 0.3~TeV, we found that they are overall consistent with the leptonic framework expectation, following the simplified argumentation in~\citet{atoyan96}. However, a direct comparison of several models to the combined \textit{Fermi}/LAT and MAGIC data suggests that they do not provide an adequate description of the Nebula flux in the energy range up to 100~TeV, yielding significant flux residuals in the 1-3~TeV energy range. This is consistent with~\cite{MAGIC_Crab15} conclusions that the IC peak is broader than expected from the modelling.

At the same time, several theoretical studies suggested that the highest energy emission of the Crab Nebula may have -- at least a partial -- contribution from the interaction of accelerated particles with the ambient medium~\citep[e.g.][]{atoyan96, bednarek97, bednarek03, amato03}. Such interactions may be intensified if particles are trapped in the Nebula filaments, leading to a $\sim 20$-fold effective target density increase~\citep{atoyan96}. The extension of the Crab Nebula synchrotron emission to $\sim 100$~MeV energies implies the presence of PeV electrons. This suggests the hadrons in the Nebula could also be accelerated to similar energies, given their generally lower losses via synchrotron radiation. Energetic protons would naturally contribute to the $\gamma$-ray flux at TeV energies via $p-p$ interactions. Several predictions from such models are plotted with dashed and dotted lines in Fig.~\ref{fig::GeV-TeV_SED}.

As can be seen from Fig.~\ref{fig::GeV-TeV_SED}, the MAGIC data disfavour a putative significant hadronic contribution to the measured TeV flux. This in turn implies that the accelerated nuclei constitute a minor fraction of the pulsar wind power and/or do not generally interact with the overdense Nebula filaments. Further VLZA observations can be used to refine this statement via a more accurate spectral shape estimation at the highest energies.

Overall, the VLZA observation technique extends MAGIC sensitivity to the highest energies, allowing us to search for galactic PeVatrons in the pre-CTA era. With an appropriate adaptation, this technique may be also used by CTA itself to further boost its sensitivity to the highest energy $\gamma$-ray events.

\begin{acknowledgements}
%
%
We would like to thank the Instituto de Astrof\'{\i}sica de Canarias for the excellent working conditions at the Observatorio del Roque de los Muchachos in La Palma. The financial support of the German BMBF and MPG, the Italian INFN and INAF, the Swiss National Fund SNF, the ERDF under the Spanish MINECO (FPA2015-69818-P, FPA2012-36668, FPA2015-68378-P, FPA2015-69210-C6-2-R, FPA2015-69210-C6-4-R, FPA2015-69210-C6-6-R, AYA2015-71042-P, AYA2016-76012-C3-1-P, ESP2015-71662-C2-2-P, FPA2017‐90566‐REDC), the Indian Department of Atomic Energy, the Japanese JSPS and MEXT, the Bulgarian Ministry of Education and Science, National RI Roadmap Project DO1-153/28.08.2018 and the Academy of Finland grant nr. 320045 is gratefully acknowledged. This work was also supported by the Spanish Centro de Excelencia ``Severo Ochoa'' SEV-2016-0588 and SEV-2015-0548, and Unidad de Excelencia ``Mar\'{\i}a de Maeztu'' MDM-2014-0369, by the Croatian Science Foundation (HrZZ) Project IP-2016-06-9782 and the University of Rijeka Project 13.12.1.3.02, by the DFG Collaborative Research Centers SFB823/C4 and SFB876/C3, the Polish National Research Centre grant UMO-2016/22/M/ST9/00382 and by the Brazilian MCTIC, CNPq and FAPERJ.\\

This research has made use of the CTA instrument response functions provided by the CTA Consortium and Observatory, see \url{http://www.cta-observatory.org/science/cta-performance/} (version prod3b-v1) for more details.
\end{acknowledgements}


\begin{appendix}
 
\section{Artificial Neural Network tools for data analysis}

Presently the multivariate analysis methods, based on artificial intelligence are extensively developed and widely used. In ground based Cherenkov gamma-ray technique, characterized with a very small level of signal events, the application of artificial neural networks (ANN) showed very good performance~\citep{Maneva2003yj}.

Both energy reconstruction and gamma-hadron classification for  VLZA analysis were checked by applying two different neural networks tools. The first one was based on JETNET Fortran package~\citep{Peterson1993nk} implemented for ROOT~\citep{ROOT} and MARS via C++ wrapper. The second one uses the modern deep learning Tensor Flow\footnote{\url{https://www.tensorflow.org/}} libraries implemented in KERAS package~\citep{chollet2015keras}. In both analyzes ANN results were added to the standard for MAGIC ROOT output files, so that the entire analysis program chain of MAGIC could be applied. We used feed-forward algorithms with back propagation minimization. The network architecture consists from an input layer, 3 hidden layers and an output layer. As input we used Hillas parameters for both telescopes as well as the results of stereoscopic reconstruction (e.g. EAS impact distance and height of the shower maximum). The performance of both tools are similar and comparable to that of the MAGIC standard~RF energy estimator. The advantage of Tensor Flow library is that it is several times faster than JETNET code, enabling better optimization of the network architecture.
 

\section{Systematic uncertainties in MAGIC VLZA measurements}
\label{sect::systematics}


The systematic uncertainties, associated with the MAGIC VLZA observations, largely overlap with those derived at lower zenith angles~\citep{Magic_performanceI, Magic_performanceII} in what concerns the telescope hardware performance. At the same time, the increased distance to EAS at large zenith angles affects the performance of the reconstruction techniques applied. Potentially, this makes them more sensitive to the otherwise small discrepancies between the MAGIC MCs and real VLZA EASs. Below we quantify the systematics contributions, specific to VLZA data taking.

\subsection{Pointing accuracy}

Increased gravitational loads during the VLZA observations, caused by close to horizontal orientation of the telescopes, result in bending of the telescope structure and camera support arch. This effect is largely compensated by an active mirror control system and contemporaneous observations of positions of a number of reference stars next to the MAGIC field of view~\citep{Magic_performanceI, Magic_performanceII}. 

To evaluate the effect from the remaining telescope mispointing, we follow the approach taken in~\citet{Magic_performanceII} and compare the true sky position of the Crab Nebula to that derived from the data on several different epochs of observations. Due to the lower count rates during the VLZA observations, such a comparison is not possible on a nightly basis. We thus combine the data in data sets spanning one or more months to properly determine the fitted source position in the skymaps. From that we determine the mispointing as the difference between the nominal and the reconstructed source positions.

\begin{figure}
	\center
	\includegraphics[width=1.0\columnwidth]{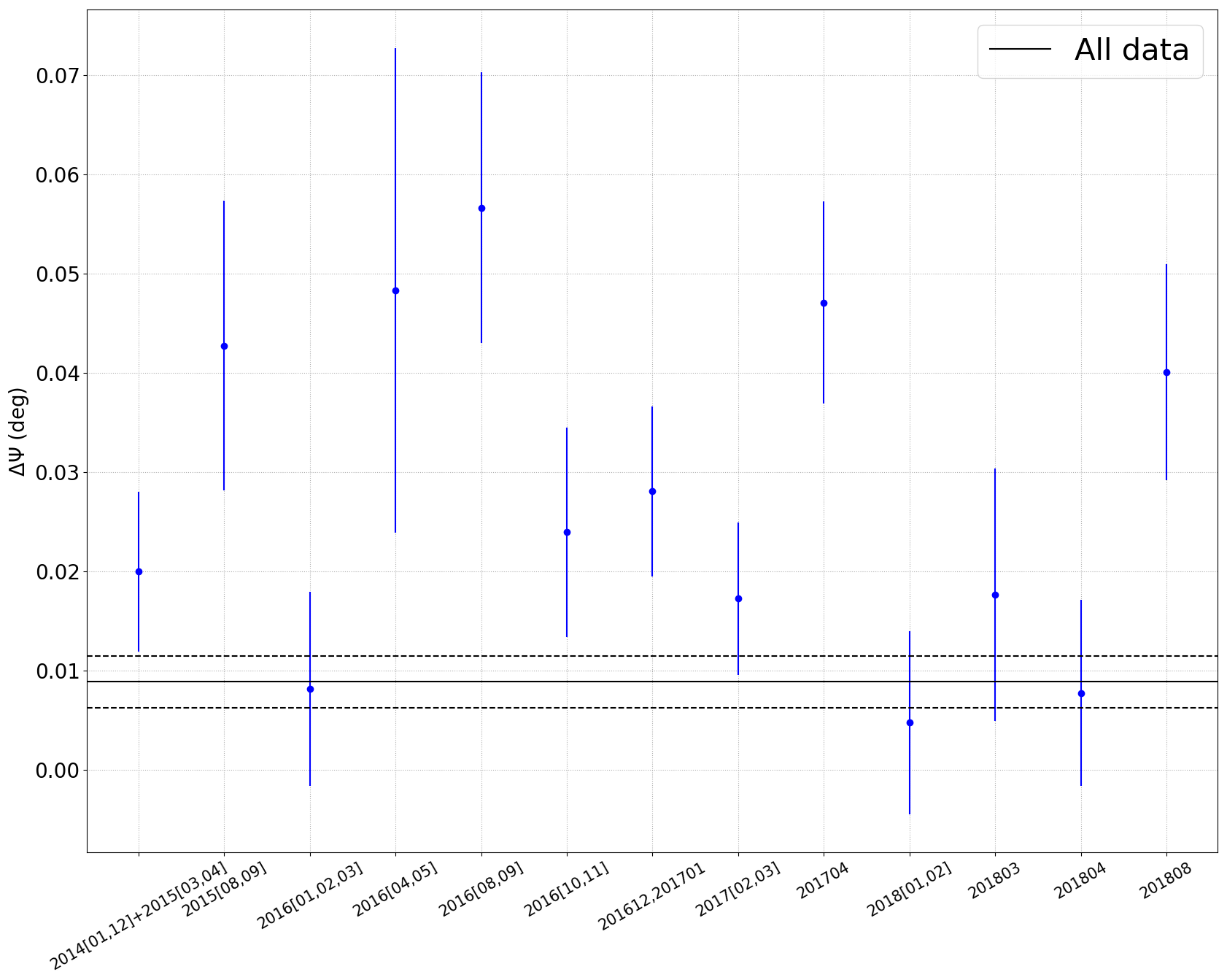}
	\caption{Time evolution of mispointing of Crab Nebula using VLZA method. Each datapoint represent one or more months of taken data. The solid line represents the results from the analysis using the complete Crab VLZA dataset, and the dashed lines represent the error on that value.  }
	\label{fig::mispointing_wrt_time}
\end{figure}

\begin{figure}
	\center
	\includegraphics[width=1.0\columnwidth]{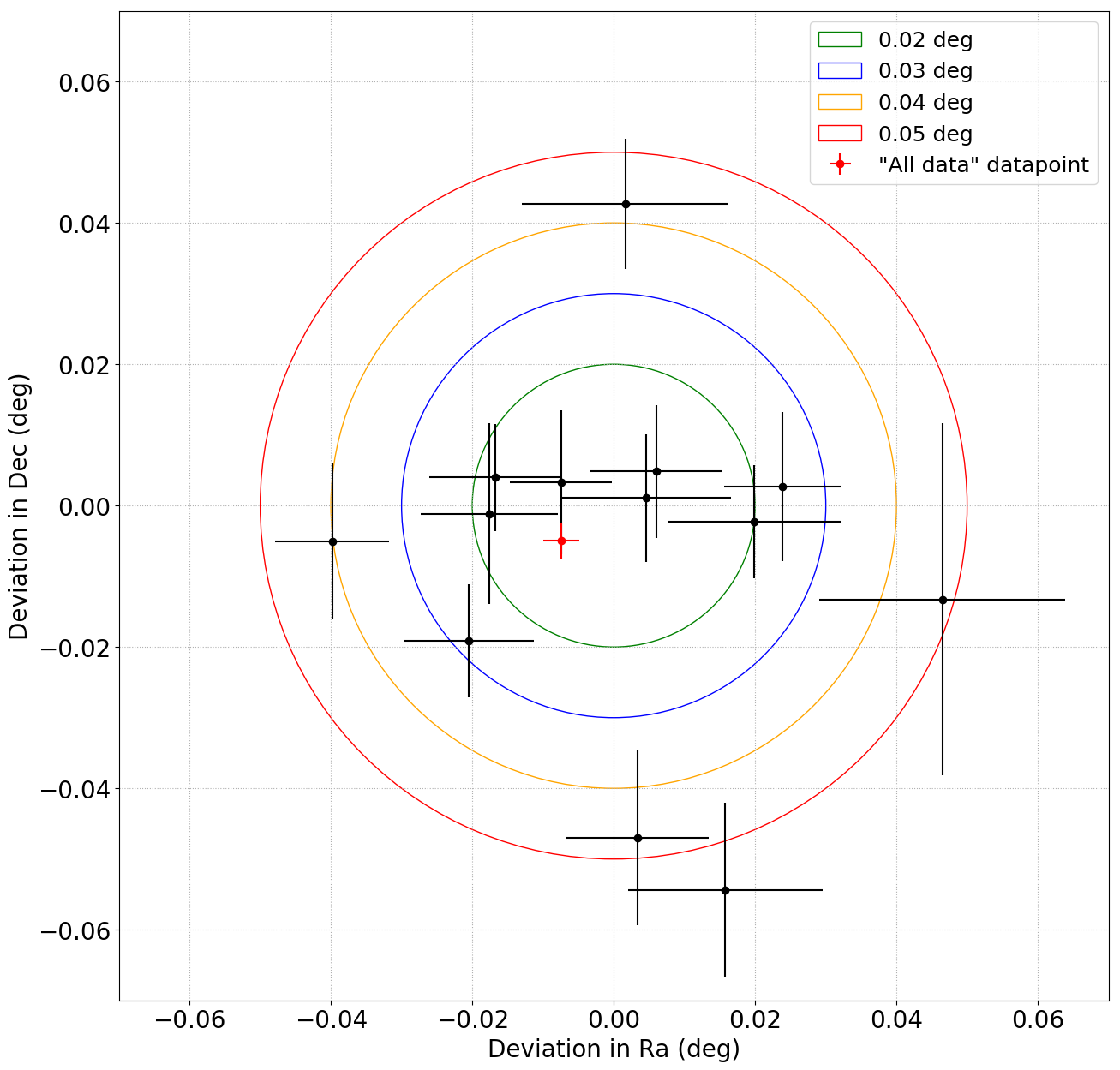}
	\caption{Mispointing in Ra/Dec of each datapoint as in Fig. \ref{fig::mispointing_wrt_time}. The circles represent specific angular distances as seen in the legend. The datapoint where all of the Crab VLZA data is analyzed is represented in red. }
	\label{fig::mispointing_2D_pull}
\end{figure}

The mispointing data versus time can be found in Fig. \ref{fig::mispointing_wrt_time}.
A two dimensional plot of the mispointings in Ra/Dec coordinates can be found in Fig. \ref{fig::mispointing_2D_pull}. We can conclude that the systematic uncertainty on the reconstructed source position is $\approx  0.04 ^\circ$. This value is larger than the one reported in ~\citet{Magic_performanceII}, which can be explained by the increased mechanical stress on the telescope structure when observing near the horizon in the VLZA regime. Note that this value is $\approx 4$ times smaller than the MAGIC point spread function (PSF) during such observations.

\begin{figure}
  \center
  \includegraphics[width=1.0\columnwidth]{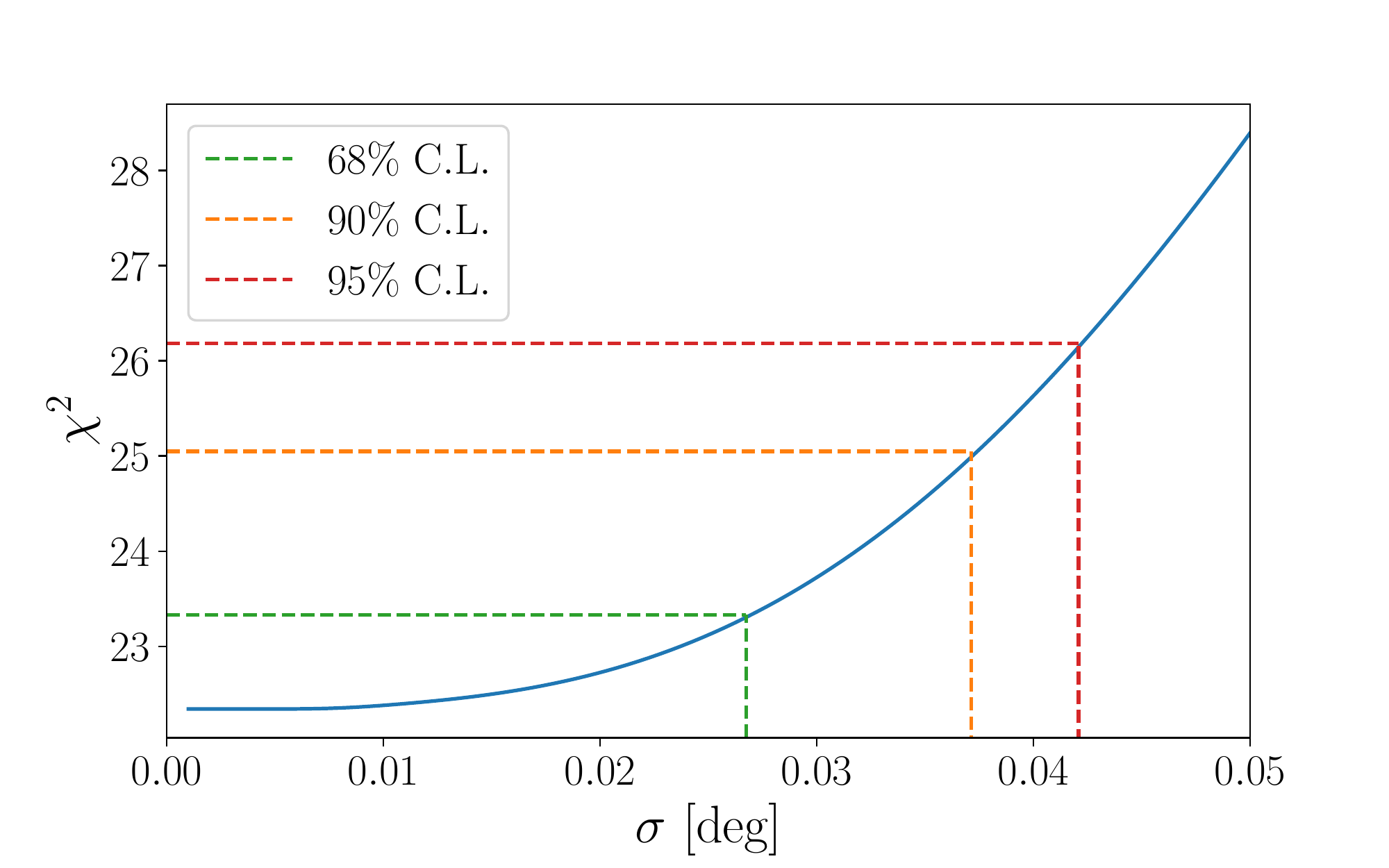}
  \caption{Scan of an additional extension of the Crab Nebula angular profile above 10~TeV photon energy on top of the MC predicted PSF. The Y axis gives the resulting $\chi^2$ value of the fit with 16 degrees of freedom. Extension is expressed by means of the $\sigma$ of the additional gaussian component. The maximal allowed extension provides the measure of the remaining MAGIC mispointing.}
  \label{fig::mispointing_with_all_data}
\end{figure}
Also, an additional estimate of the remaining systematic uncertainty on the event direction reconstruction can be obtained from the extension of the reconstructed Crab Nebula total angular profile with respect to the Monte Carlo estimated PSF. The MAGIC PSF shape is well described by a King function~\citep{Magic_performanceII, DaVela_PSF}, whereas the putative additional mispointing random in its nature would appear as an additional smearing of this profile. The fit of the VLZA Crab Nebula data above 10~TeV suggests that such additional mispointing does not exceed $0.04^\circ$ at 90\% confidence level, as shown in Fig.~\ref{fig::mispointing_with_all_data}.

The additional mispointing yields a broader PSF than predicted by Monte Carlo simulation, thus affecting the true event containment within a given off-source angle cut. The resulting effect depends on the original (energy dependent) PSF width and thus changes with the energy. Using the MAGIC VLZA simulations we estimate the impact of this mispointing to be $\lesssim 4\%$ at $\sim 3$~TeV and $\lesssim 8\%$ at $\sim 30$~TeV energies.


\subsection{Zenith angle dependence of the instrument response}

\begin{figure}
  \center
  \includegraphics[width=1.0\columnwidth]{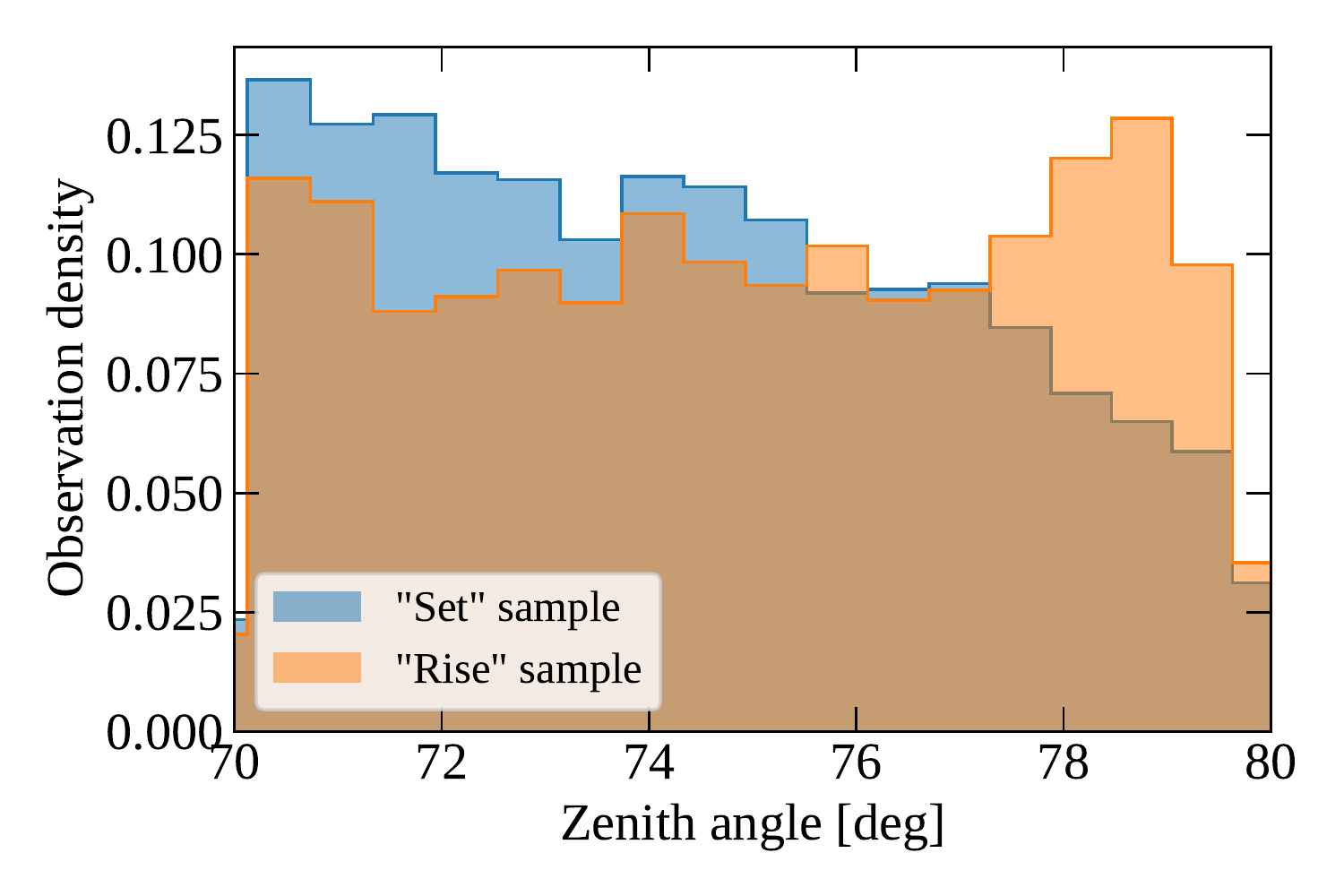}
  \caption{Zenith angle distribution of the accumulated VLZA Crab Nebula observational sample, used to re-weight MAGIC response functions.}
  \label{fig::obs_zd_distribution}
\end{figure}

The rapid growth of the air mass with the zenith angle in the $70\deg - 80\deg$ range results in a gradual change of the MAGIC response with respect to the incoming EAS. To account for this effect we split the MAGIC MC sample into 100 bins in cosine of zenith angle (in the $0\deg - 90\deg$ range; such binning roughly follows the growth of the air mass). The instrument response functions were then computed re-weighting this sample with the zenith angle distribution of the accumulated data, shown in Fig.~\ref{fig::obs_zd_distribution}.

\subsection{MC to data comparison}

EAS development at zenith angles above $70^\circ$ proceeds primarily in the rarefied layers of the upper atmosphere and at $\gtrsim 50-100$~km distances from the observer. These conditions lead to certain peculiarities in the shower evolution~\citep[dependent on the nature of the primary particle, see][]{neronov16}. Due to this,
VLZA observations may be associated with a larger MC to data discrepancy, compared to that derived from lower zenith angle data~\cite{Magic_performanceII}.

\begin{figure*}
  \center
  \includegraphics[width=1.0\textwidth]{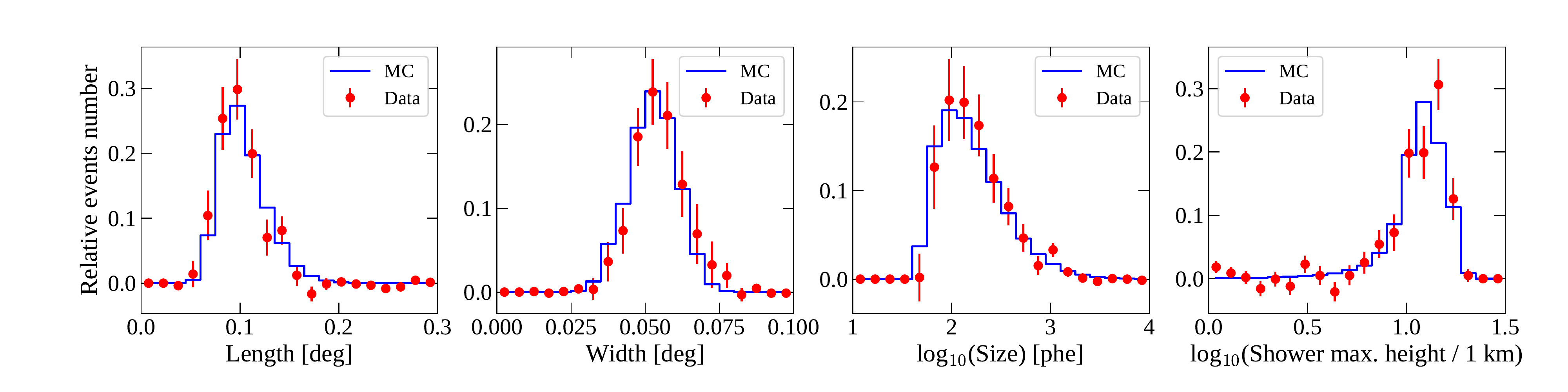}
  \caption{Comparison of basic EAS reconstruction parameters between the MC simulated (blue) and real (red) event samples, recorded in the 70-75 deg zenith angle range. \textit{Size}, \textit{Length} and \textit{Width} are the so-called Hillas parameters~\citep{hillas85}, whereas the shower maximal height is reconstructed from the MAGIC stereo data by a standard analysis pipeline in MARS. Deviation of real distributions from MC does not exceed the $2\sigma$~confidence level.}
  \label{fig::mc-data_comparison}
\end{figure*}
In order to verify this, we have compared the distribution of the basic EAS ``Hillas'' parameters \textit{Size}, \textit{Length}, \textit{Width}~\citep{hillas85} as well as the height of the maximal shower development in MAGIC Monte Carlo and real event samples. For the latter we have used the excess distributions of the same parameters in the on- and off-source regions, derived with loose event selection cuts. This comparison is shown in Fig.~\ref{fig::mc-data_comparison} for events in the zenith angle range 70-75 deg where no significant difference is present between the real and simulated Crab Nebula VLZA signals.


\subsection{Energy bias and resolution}

We estimate the MAGIC energy resolution and bias comparing the true MC simulated event energies to those obtained with our energy reconstruction algorithms. In order to parameterise the accuracy of the reconstruction we fit a gaussian to the scaled $(E_{est} - E_{true}) / E_{true}$ distribution of the estimated energies $E_{est}$ in narrow bins of $E_{true}$. The mean of this distribution is taken as a measure of bias, whereas its width corresponds to the energy resolution of the applied reconstruction procedure.

\begin{figure*}
  \center
  \includegraphics[width=0.9\textwidth]{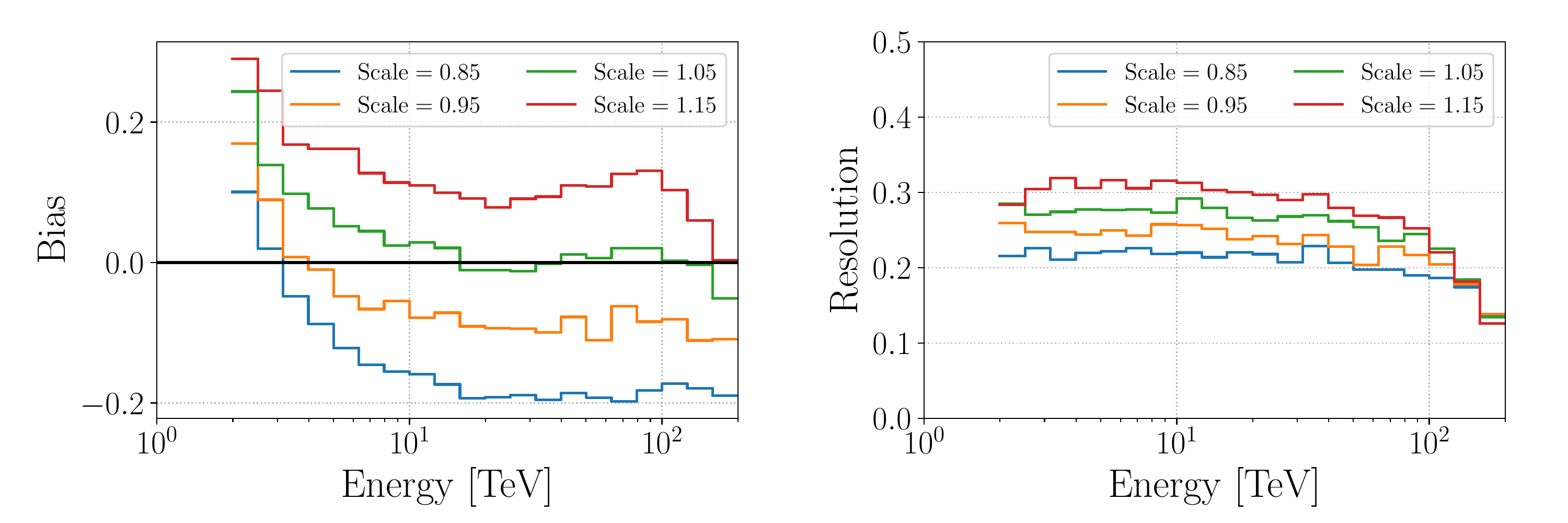}
  \caption{
    \textit{Left:} bias in the VLZA energy estimation, computed for different scaling values of the MAGIC light throughput system. It is estimated as the mean of the gaussian distribution of the estimated MC energies and is given relative to the assumed true energy in the simulation.
    \textit{Right:} VLZA relative energy resolution (for different scaling values of the MAGIC light throughput system), computed as the sigma of the gaussian distribution of the estimated MC energies.
  }
  \label{fig::energy_resolution_and_bias}
\end{figure*}

The resulting energy estimate is subject to uncertainties in the overall MAGIC light throughput, which are estimated to be $\lesssim 15\%$~\citep{Magic_performanceII}. To estimate the possible impact on the VLZA energy reconstruction, we apply an additional scaling of the amount of light in the VLZA MC sample by $\pm 15\%$. These ``scaled'' MCs are then processed as if no light scale was applied. The bias and resolution resulting from them (as a function of the true event energy) are given in Fig.~\ref{fig::energy_resolution_and_bias}. 

As it can be seen from there, the overall energy bias varies in the range $[-20\%;+15\%]$, which gives an estimate of the instrumental MAGIC energy scale uncertainty in the VLZA regime. Since the total amount of light recorded from EAS plays a major role in the event energy reconstruction, it is worth to note that the resulting energy bias is almost directly proportional to the assumed light scale.

\subsection{Atmospheric transmission}
\label{sect::sbig_correction}

Atmospheric transmission directly affects the amount of EAS light, reaching the telescope camera. The uncertainties on it contribute to the overall uncertainty of the VLZA energy scale, as discussed above.

To estimate the wavelength-dependent atmospheric transmission, we image the stellar field next to Crab Nebula with dedicated CCD cameras, located close to the centers of the MAGIC-I and MAGIC-II reflector dishes. The images were taken every 90 seconds with the 90 second exposure, cyclically changing the colour filters from red ($\lambda_{mean} \sim 640$~nm) to green ($\lambda_{mean} \sim 530$~nm) and to blue ($\lambda_{mean} \sim 450$~nm). The acquired images were flat-fielded and cleaned of hot pixels and dark current contribution. Then counts from selected bright stars were estimated as a difference of counts from the circular region around the star and the background counts from an annular region of a larger diameter.

In order to calibrate this aperture photometry procedure, an additional imaging of this stellar field was performed on several clean nights, when atmosphere absorption showed no variation with height. During such nights light absorption in each colour filter follows a simple law:
\begin{equation}
  c = c_{0} \exp{ (-\alpha m_{air}(z)) }
  \label{eq::SBIG_light_absorption}
\end{equation}
where $c$ is the number of background-subtracted CCD counts, $c_0$ is the number of counts before absorption, $\alpha \approx {\rm const}$ is the specific absorption coefficient and $m_{air}$ is the air mass at a given zenith angle $z$. 
The constant $c_0$ can be determined from Eq.~\ref{eq::SBIG_light_absorption} from measured CCD counts at different zenith angles from the selected star. Knowing $c_0$, the average absorption coefficient $\alpha$ during the subsequent observational sessions can be estimated as $\alpha = -\log{(c/c_0)}/m_{air}(z)$.

Contemporaneous imaging of those selected stars during the VLZA data taking allows to estimate atmospheric transmission for EAS with temporal resolution of $1.5-3$~minutes. The height of each shower maximum, estimated as a part of the standard data reduction in MARS, is used to compute the line of sight distance to the shower maximum and derive the corresponding value of the air mass $m_{air}^{EAS}$. The resulting absorption then can be estimated as $\tau_{data} = \exp{ (-\alpha m_{air}^{EAS}(z)) }$. The ratio of this latter value to the absorption assumed in the MAGIC detector Monte Carlo simulations (for the same zenith angle and shower distance) defines the relative light scale $s = \tau_{data} / \tau_{MC}$ in each of the colour filters, which -- after a convolution with the Cherenkov emission spectrum -- is finally used to correct the estimated event energies or select time intervals with good transmission.

Though the stellar light measurements, described above, provide a simple and reliable way to estimate the total atmospheric transmission, they are subject to inaccuracies due to the uncertainties in the derived calibration constants $c_{0}$ and uncertainties in the measured CCD counts during the observations. We have minimized the latter by choosing the camera exposure time so that the reference stars get $\gtrsim 3 \times 10^4$ CCD counts, so that the resulting uncertainty is less than 1\%.

\begin{figure*}
  \center
  \includegraphics[width=0.9\columnwidth]{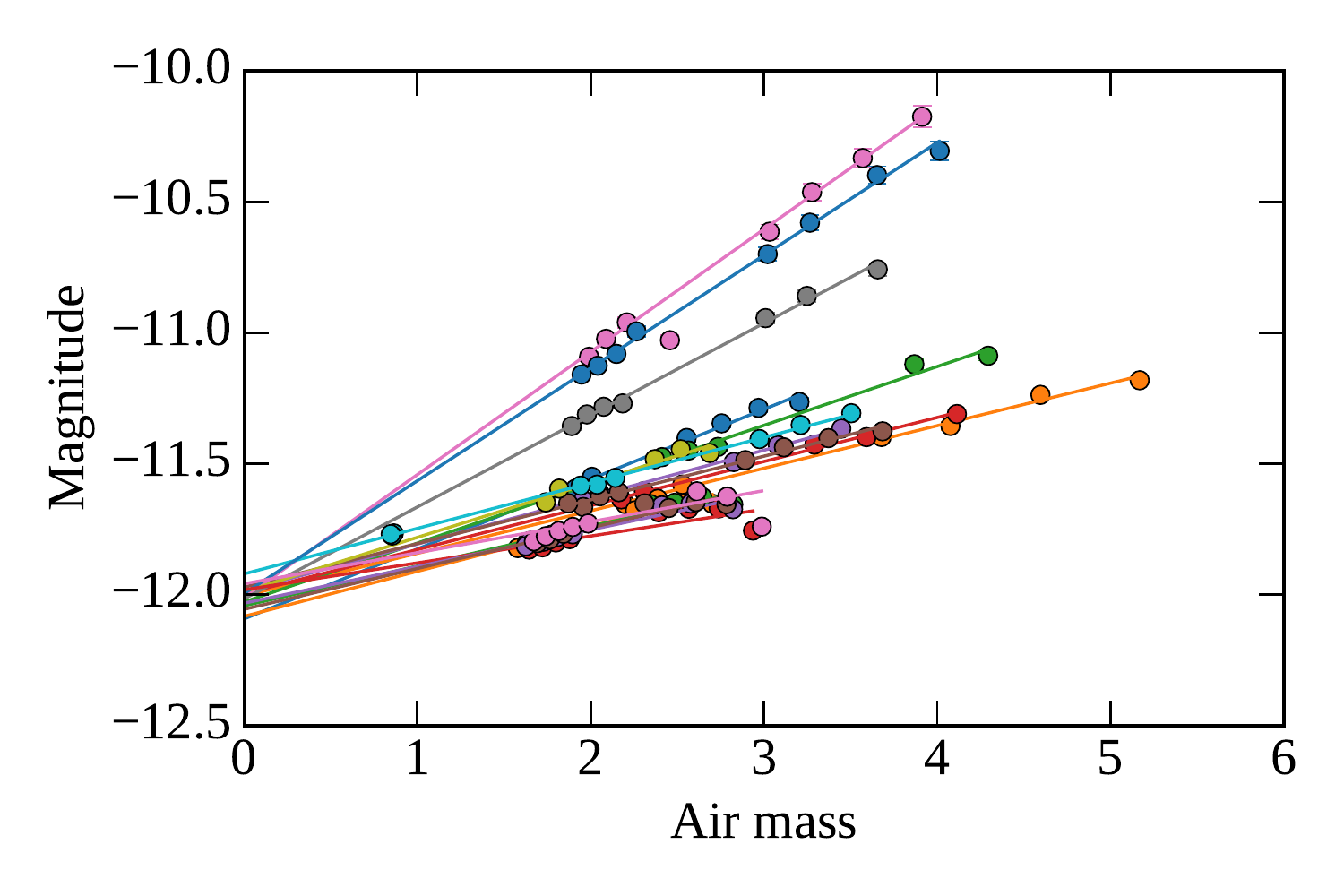}
  \caption{Recorded CCD counts $c$ from the star HD35708 as a function of air mass, converted to the stellar magnitudes $m = m_0 - \log_{2.512} c$ with an arbitrary reference point $m_0$. These counts were obtained with the green filter during the CCD imaging of the Crab Nebula region on nights with stable atmosphere transparency. Solid lines correspond to the each night fits of Eq.~\ref{eq::SBIG_light_absorption} assuming $\alpha = {\rm const}$; their extrapolation to the zero air mass defines the reference calibration constant $c_0$ for the particular filter used. The RMS of these extrapolations sets the $c_0$ uncertainty and is $\approx 0.04$~mag for this calibration sample.
  }
  \label{fig::sbig_calibration}
\end{figure*}

The uncertainty on the calibration constants $c_{0}$ was computed from several $c_{0}'$ estimates, taken on nights with stable atmosphere transparency ($<10\%$ deviations from the Eq.~\ref{eq::SBIG_light_absorption} law), as shown in Fig.~\ref{fig::sbig_calibration}. The standard deviation of these estimates suggests that the calibration constants for the reference stars are determined with the accuracy $\lesssim 5\%$.

The calculation of the exact transmission correction for a specific air shower from such stellar light measurements depends on the assumed distribution of the absorber in the atmosphere, which induces an additional uncertainty. Our estimates suggest that, though for mild ($\lesssim 20\%$) light absorption the transmission estimates are generally accurate to within 3-5\%, deviations up to $10\%$ are still possible in some cases. We thus conservatively use this latter value as an estimate of the atmosphere transmission systematics.


\subsection{Total systematic uncertainty}

To estimate the total systematic uncertainty in the VLZA case, we also account to non-VLZA specific sources of MAGIC systematics, reviewed in~\citet{2012MAGICPerformance}; the resulting list of contributions is given in Tab.~\ref{tab::sys_summary}. Combining these we find that the telescope's energy scale is determined with the accuracy of~19\% at low ($\sim 3$~TeV) and~17\% at medium ($\sim 30$~TeV) energies. This is comparable to the MAGIC energy scale systematics estimated from the muon analysis, worsened by the larger uncertainty in the atmosphere transmission due to VLZA conditions. The uncertainty on the reconstructed flux normalization (excluding the effect of the energy scale) is~14\% and~20\% correspondingly.


\begin{table}
    \centering
    \caption{Summary of the MAGIC Crab Nebula VLZA observations systematics. The values not specific to the VLZA data case are taken from~\citet{2012MAGICPerformance}. The values affecting the telescope energy scale and flux normalization are marked with ``ES'' and ``FN'' correspondingly.}
    \begin{tabular}{l|l}
        \hline\hline
        Systematic effect                  & Resulting uncertainty   \\
        \hline
        F-Factor                           & 10\% ES                 \\
        Atmospheric transmission           & <10\% ES                \\
        Mirror reflectivity                & 8\% ES                  \\
        PMT electron collection efficiency & 5\% ES                  \\
        Light collection in a Winston Cone & 5\% ES                  \\
        PMT quantum efficiency             & 4\% ES                  \\
        Signal extraction                  & 3\% ES                  \\
        Temperature dependence of gains    & 2\% ES                  \\
        Charge flat-fielding               & 2-8\% ES FN             \\
        Analysis and MC discrepancies      & <10-18\% FN             \\
        Background subtraction             & 2\% FN                  \\
        Broken channels/pixels             & 3\% FN                  \\
        Mispointing                        & 4-8\% FN                \\
        NSB                                & 1-4\% FN                \\
        Trigger                            & 1\% FN                  \\
        Unfolding of energy spectra        & 0.1 SL                  \\
        \hline
    \end{tabular}
    \label{tab::sys_summary}
\end{table}

\end{appendix}

\bibliographystyle{aa}
\bibliography{References}

\end{document}